\documentclass[aps,prb,amsmath,amssymb,twocolumn,floatfix]{revtex4}
\usepackage{graphicx}
\usepackage{amssymb}
\usepackage{amsmath}
\usepackage{epstopdf}
\usepackage{subfigure}
\usepackage[usenames,dvipsnames]{color}   
\usepackage{hyperref}
\usepackage{placeins}
\usepackage{sidecap}
\usepackage{color}
\makeatletter

\newcommand{\bra}[1]{\langle #1 |}
\newcommand{\ket}[1]{| #1 \rangle}
\newcommand{\braket}[2]{\langle #1 | #2 \rangle}

\newcommand{\elecspin}[1]{\hat {\bf S}_#1}

\begin{document}
\title{Floquet theory of radical pairs in radiofrequency magnetic fields}
\author{Hamish G. Hiscock, Daniel R. Kattnig, David E. Manolopoulos and P. J. Hore}
\affiliation {Department of Chemistry, Physical and Theoretical Chemistry Laboratory, 
University of Oxford, South Parks Road, Oxford OX1 3QZ, United Kingdom}

\begin{abstract}
\noindent We present a new method for calculating the product yield of a radical pair recombination reaction in the presence of a weak time-dependent magnetic field. This method successfully circumvents the computational difficulties presented by a direct solution of the Liouville-von Neumann equation for a long-lived radical pair containing many hyperfine-coupled nuclear spins. Using a modified formulation of Floquet theory, treating the time-dependent magnetic field as a perturbation, and exploiting the slow radical pair recombination, we show that one can obtain a good approximation to the product yield by considering only nearly-degenerate sub-spaces of the Floquet space. Within a significant parameter range, the resulting method is found to give product yields in good agreement with exact quantum mechanical results for a variety of simple model radical pairs. Moreover it is considerably more efficient than the exact calculation, and it can be applied to radical pairs containing significantly more nuclear spins. This promises to open the door to realistic theoretical investigations of the effect of radiofrequency electromagnetic radiation on the  photochemically induced radical pair recombination reactions in the avian retina which are believed to be responsible for the magnetic compass sense of migratory birds.
\end{abstract}

\maketitle

\section{Introduction}

A radical pair (RP) is a pair of transient radicals whose unpaired electron spins, one on each radical, are correlated. Provided the RP is formed in a spin conserving process, such as photo-induced electron transfer in a precursor molecule, the initial spin state is known and this information evolves over time under the influence of external magnetic fields (Zeeman interactions) as well as internal electron-electron (exchange and dipolar) and electron-nuclear hyperfine coupling. In the simplest reaction scheme, the RP recombines to different products depending on the electronic spin state -- singlet, S or triplet, T (Fig.~1). The relative yields of the two products therefore contain information about the interactions experienced by the RP during its lifetime, including a dependence on the direction of the external magnetic field. This is the physical basis of the proposal that RP reactions are responsible for the magnetic compass sense of migratory birds.\cite{Schulten78,Hore16}

There is a growing body of evidence in support of this proposal. Behavioural studies have shown that migratory birds are only magnetically oriented in the presence of blue/green light.\cite{Wiltschko99,Muheim02,Wiltschko10} This is broadly consistent with the photo-excitation of the flavin adenine dinucleotide (FAD) cofactors in cryptochrome proteins in the avian retina which are believed to be the precursors of the RP reactions.\cite{Ritz00,Maeda12} Studies have also shown that the avian magnetic sensor is an inclination compass rather than a polarity compass: it is insensitive to inverting the direction of the magnetic field.\cite{Wiltschko72,Wiltschko95} This is consistent with time-reversal symmetry in the RP mechanism. And more recent studies have found that birds are disoriented when exposed to weak radiofrequency (RF) electromagnetic fields.\cite{Ritz04,Thalau05,Ritz09,Engels14,Kavokin14,Wiltschko15,Schwarze16} If this were properly understood, it might perhaps provide the most compelling evidence of all for RP-based magnetoreception.\cite{Hore16} However, theoretical calculations of the effect of weak RF magnetic fields on the RP mechanism are currently limited, and the existing experimental evidence is still somewhat controversial.

The effect of RF radiation on avian magnetoreception was first observed in a study by Ritz {\em et al.}\cite{Ritz04} in 2004, in which both broadband noise and single frequency electromagnetic fields were found to disrupt the ability of birds to magnetically orient. Following this initial finding, there have been a number of further experimental studies aimed at characterising the interaction between the RF magnetic field and the magnetosensor.\cite{Thalau05,Ritz09,Engels14,Kavokin14,Wiltschko15,Schwarze16} One of the most influential of these found that irradiation at the ``Larmor" frequency -- the precession frequency of a free electron spin in the Earth's magnetic field {(1.4 MHz at 50 $\mu$T)} -- had a special resonance effect, causing disorientation at even very weak intensity levels ($\ge 15$ nT).\cite{Ritz09} If this were correct, it would imply that the electron spin on one of the radicals in the RP magnetosensor is coupled only to the external field, and not to any nuclear spins. This in turn would impose a stringent constraint on the identity of the RP, as there are few radicals without any nuclear spins in biological systems. However, a replication study has since failed to show the same sensitivity to Larmor frequency perturbations, and found instead that broadband noise is far more effective at disrupting the avian compass.\cite{Engels14,Schwarze16}

Given these conflicting experimental results, a theoretical study of how the time-dependent magnetic field component of RF radiation affects the RP spin dynamics would clearly be very useful. However, there are several obstacles to such a study. Firstly, at least one (and quite possibly both) of the biological radicals involved is likely to have many nuclear spins with anisotropic hyperfine interactions. Since the size of the Hilbert space increases exponentially with the number of spins, even simulations involving a static external field are computationally expensive. Secondly, if we wish to treat broadband noise perturbations, the time dependence of the external field will not be monochromatic as is required for many existing approximate methods.\cite{Timmel96,Rodgers05,Rodgers09} Additionally, as noted by Gauger {\em et al.},\cite{Gauger11} for a RP to exhibit the observed sensitivity to weak time-dependent fields, its lifetime must be exceedingly long. 

{One exact method that is applicable to the problem is the numerical solution of the Liouville-von Neumann equation, but this becomes cripplingly computationally expensive for long simulation times and realistic RPs. An alternative is provided by the COMPUTE algorithm,\cite{Compute1,Compute2} which addresses the need for a long simulation time by constructing the propagator for a full modulation period of the (periodic) RF field. We have considered using this algorithm for the present problem, but found that a large number of matrix exponentials are needed to construct the propagator when the RF field contains many Fourier components, and that the expense of evaluating these exponentials becomes prohibitive when the RP contains many nuclear spins.}

We therefore require a new approximate method capable of treating large spin systems in the presence of weak external magnetic fields with a complex time dependence (involving many Fourier components). In the following sections, we shall first present and then validate such a method, based on a modified formulation of Floquet theory. {Various versions of this theory are well known and widely used in fields ranging from solid state nuclear magnetic resonance to multiphoton spectroscopy.\cite{Scholz10,Leskes10,Chu04,Eckardt15} However, we are not aware of any previous application of Floquet theory to radical pair recombination reactions in time-dependent magnetic fields.}

\begin{figure}[t]
\begin{center}
 \includegraphics[scale=0.25]{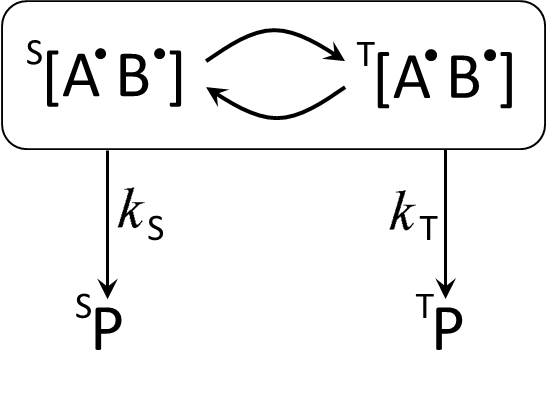}
 \caption{A schematic illustration of a RP recombination reaction. Once the radical pair has formed, its singlet and triplet states $^{\rm S}[{\rm A}^{\boldsymbol{\cdot}}{\rm B}^{\boldsymbol{\cdot}}]$ and $^{\rm T}[{\rm A}^{\boldsymbol{\cdot}}{\rm B}^{\boldsymbol{\cdot}}]$ interconvert by hyperfine- and Zeeman-mediated intersystem crossing, and recombine to give distinct products with rate constants $k_{\rm S}$ and $k_{\rm T}$.}
\end{center}
\end{figure}

\section{Floquet Theory}

\subsection{The propagator}

We are interested in treating a quantum system evolving under a Hamiltonian which is periodic in time with period $T=2\pi/\omega$,
\begin{equation}
 \hat H(t) = \sum_{n=-\infty}^{\infty} \hat{H}^{(n)} e^{+in\omega t}.
\end{equation}
Shirley\cite{Shirley65} was the first to employ Floquet's theorem\cite{Magnus79} to show that this is equivalent to solving an infinite dimensional eigenvalue problem involving a so-called Floquet Hamiltonian with matrix elements
\begin{equation}
 \bra{\beta n}\hat H_{\rm F}\ket{\alpha m} = H_{\beta\alpha}^{(n-m)} + n\omega\,\delta_{nm}\delta_{\beta\alpha}.
\end{equation}
Here $\ket{\alpha m}$ is a basis state in an extended ``Floquet space", the product of a Hilbert space basis state $\ket{\alpha}$ and a Fourier space basis state $\ket{m}$ with $\braket{t}{m}=e^{+im\omega t}/\sqrt{2\pi}$, and 
\begin{equation}
H_{\beta\alpha}^{(n-m)} = \bra{\beta}\hat{H}^{(n-m)}\ket{\alpha}.
\end{equation}
In particular, Shirley showed that the matrix element of the evolution operator between states $\ket{\alpha}$ and $\ket{\beta}$ in Hilbert space can be written as\cite{Shirley65}
\begin{equation}
  U_{\beta\alpha}(t;t_0) = \sum_{n}\bra{\beta n}e^{-i\hat H_{\rm F}(t-t_0)}\ket{\alpha 0} e^{+in\omega t}.
\end{equation}

The eigenvalues $\varepsilon_{\gamma l}$ and eigenstates $\ket{\varepsilon_{\gamma l}}$ of the Floquet Hamiltonian exhibit certain periodicity properties, which are endowed by the structure of $\hat H_{\rm F}$,\cite{Shirley65}
\begin{equation}
 \varepsilon_{\gamma\,l+p} = \varepsilon_{\gamma l} + p\,\omega,
\end{equation}
\begin{equation}
 \braket{\alpha\, n+p}{\varepsilon_{\gamma\, l+p}} = \braket{\alpha\, n}{\varepsilon_{\gamma l}}.
\end{equation}
These relations are in fact required to ensure the unitarity of the propagator in Eq.~(4), and they can be used to re-write $U_{\beta\alpha}(t;t_0)$ as follows:\cite{Charpentier98}
\begin{equation}
\begin{aligned}
&U_{\beta\alpha}(t;t_0) = \sum_{n\gamma l} \braket{\beta n}{\varepsilon_{\gamma l}}e^{-i\varepsilon_{\gamma l}(t-t_0)}\braket{\varepsilon_{\gamma l}}{\alpha 0}e^{+in\omega t}\\
&= \sum_{n\gamma l} \braket{\beta 0}{\varepsilon_{\gamma\, l-n}}e^{-i\varepsilon_{\gamma\, l-n}(t-t_0)}
\braket{\varepsilon_{\gamma\, l-n}}{\alpha (-n)}e^{+in\omega t_0}\\
&= \sum_{n\gamma m} \braket{\beta 0}{\varepsilon_{\gamma m}}e^{-i\varepsilon_{\gamma m}(t-t_0)}\braket{\varepsilon_{\gamma m}}{\alpha (-n)}e^{-i(-n) \omega t_0}\\
&=\sum_{n}\bra{\beta 0}e^{-i\hat H_{\rm F}(t-t_0)}\ket{\alpha n} e^{-in\omega t_0}.
\end{aligned}
\end{equation}

\subsection{Expectation values}

Consider the expectation value of an observable $\langle A(t;t_0)\rangle$ at time $t$ subject to some initial condition at time $t_0$. In the density operator formulation, this is 
\begin{equation}
 \langle A(t;t_0)\rangle = {\rm tr}\left[\hat A\, \hat{U}(t;t_0) \hat \rho(t_0) \hat{U}(t;t_0)^\dagger\right],
\end{equation}
where $\hat \rho(t_0)$ is the density operator at time $t_0$. Inserting resolutions of the identity in Hilbert space, this becomes
\begin{equation}
 \langle A(t;t_0)\rangle = \sum_{\alpha\beta\gamma\delta} A_{\alpha\beta} 
 U_{\beta\gamma}(t;t_0)\rho_{\gamma\delta}(t_0)U_{\alpha\delta}(t;t_0)^*,
\end{equation}
where $A_{\alpha\beta}=\bra{\alpha}\hat{A}\ket{\beta}$ and $\rho_{\gamma\delta}(t_0)=\bra{\gamma}\hat{\rho}(t_0)\ket{\delta}$. If we use Eq.~(7) for the matrix elements of the time evolution operator, we obtain
\begin{equation}
\begin{aligned}
\langle A(t;t_0)\rangle &= \sum_{\alpha\beta\gamma\delta}\sum_{mn} A_{\alpha\beta}
\bra{\beta 0}e^{-i\hat H_{\rm F}(t-t_0)}\ket{\gamma m}e^{-im\omega t_0}\\
&\times \rho_{\gamma\delta}(t_0)e^{+in\omega t_0}\bra{\delta n}
e^{+i\hat H_{\rm F}(t-t_0)}\ket{\alpha 0}.
\end{aligned}
\end{equation}
Introducing a trace over Floquet space, this can be written in a form analogous to Eq.~(8),
\begin{equation}
 \langle A(t;t_0)\rangle = {\rm tr}_{\rm F}\left[\hat A_{\rm F}\,e^{-i\hat H_{\rm F}(t-t_0)}\hat \rho_{\rm F}(t_0)\, e^{+i\hat H_{\rm F}(t-t_0)}\right],
\end{equation}
where the Floquet space detection and density operators are
\begin{equation}
\hat A_{\rm F} = \sum_{\alpha\beta} \ket{\alpha 0}A_{\alpha\beta}\bra{\beta 0},
\end{equation}
\begin{equation}
 \hat \rho_{\rm F}(t_0) = \sum_{\alpha\beta}\sum_{mn}\ket{\alpha m}e^{-im\omega t_0}\rho_{\alpha\beta}(t_0)e^{+in\omega t_0}\bra{\beta n}.
\end{equation}

Now had we used Eq.~(4) for $U_{\beta\alpha}(t;t_0)$ rather than Eq.~(7), we would have obtained the formulation of Floquet theory described by Ernst and co-workers,\cite{Levante95} in which the Floquet space detection and density operators are
\begin{equation}
\hat A_{\cal F}(t) = \sum_{\alpha\beta}\sum_{mn} \ket{\alpha m}e^{+im\omega t}A_{\alpha\beta}\,e^{-in\omega t}\bra{\beta n},
\end{equation}
\begin{equation}
 \hat \rho_{\cal F}(t_0) = \sum_{\alpha\beta}\ket{\alpha 0}\rho_{\alpha\beta}(t_0)\bra{\beta 0}.
\end{equation}
However, this is less convenient for our purposes, because the detection operator $\hat{A}_{\cal F}(t)$ contains time-dependent phase factors. These lead to less convenient time-independent expressions for observables such as the singlet yield of a radical pair recombination reaction that we shall consider below.

Our initial density operator $\hat{\rho}_{\rm F}(t_0)$ also contains phase factors, but these are easier to deal with, because they are evaluated at time $t_0$ rather than time $t$. Indeed it is often natural to specify the initial conditions at time $t_0=0$, in which case the phase factors in $\hat{\rho}_{\rm F}(t_0)$ disappear and Eqs.~(11) and (13) simplify to
\begin{equation}
 \langle A(t)\rangle = {\rm tr}_{\rm F}\left[\hat A_{\rm F}\,e^{-i\hat H_{\rm F}t}\hat \rho_{\rm F}(0)\, e^{+i\hat H_{\rm F}t}\right],
\end{equation}
\begin{equation}
 \hat \rho_{\rm F}(0) = \sum_{\alpha\beta}\sum_{mn}\ket{\alpha m}\rho_{\alpha\beta}(0)\bra{\beta n}.
\end{equation}
If the eigenspectrum of $\hat{H}_{\rm F}$ can be found or approximated, Eq.~(16) can be evaluated by inserting resolutions of the identity in Floquet space,
\begin{equation}
\begin{aligned}
\langle A(t)\rangle &= \sum_{\alpha\beta} \sum_{mn} \bra{\varepsilon_{\alpha m}}\hat A_{\rm F}\ket{\varepsilon_{\beta n}}
\bra{\varepsilon_{\beta n}}\hat \rho_{\rm F}(0)\ket{\varepsilon_{\alpha m}}\\
&\times e^{+i (\varepsilon_{\alpha m} - \varepsilon_{\beta n})t}.
\end{aligned}
\end{equation}

\subsection{Matrix representations}

Before we move on to discuss the diagonalisation of $\hat{H}_{\rm F}$, it might be useful to summarise the various operators that appear in Eq.~(16) in matrix notation:
\begin{equation}
 H_{\rm F} =
\left(\begin{array}{c|c|c|c|c}
... & & & &... \\ \hline
 &H^{(0)} + \omega I &H^{(1)} &H^{(2)} & \\ \hline
 &H^{(-1)} &H^{(0)} &H^{(1)} & \\ \hline
 &H^{(-2)} &H^{(-1)} &H^{(0)} - \omega I & \\ \hline
... & & & &... \\
\end{array}\right),
\end{equation}
\begin{equation}
 A_{\rm F} = \left(\begin{array}{c|c|c|c|c}
... & & & &... \\ \hline
 &0 &0 &0 & \\ \hline
 &0 &A &0 & \\ \hline
 &0 &0 &0 & \\ \hline
... & & & &... \\
\end{array}\right),
\end{equation}
\begin{equation}
  \rho_{\rm F}(0) = \left(\begin{array}{c|c|c|c|c}
... & & & &... \\ \hline
 &\rho(0) &\rho(0) &\rho(0) & \\ \hline
 &\rho(0) &\rho(0) &\rho(0) & \\ \hline
 &\rho(0) &\rho(0) &\rho(0) & \\ \hline
... & & & &... \\
\end{array}\right).
\end{equation}
Here each individual block (separated by horizontal and vertical lines) is the size of the Hilbert space, and $I$ is the Hilbert space identity matrix.

\section{Perturbation treatment}

The evaluation of Eq.~(18) requires the solution of a {\em time-independent} matrix problem: the diagonalisation of $H_{\rm F}$ and the transformation of $A_{\rm F}$ and $\rho_{\rm F}(0)$ into its eigenstate basis. Clearly, because the matrices involved are infinite dimensional, this can only be done  approximately.

\subsection{Diagonalising $H_{\rm F}$}

One way to tackle the problem is to truncate the Floquet Hamiltonian, including a certain number of Hilbert space sized blocks and diagonalising this enlarged matrix numerically.\cite{Maquet83} This method may offer a practical solution when there are only one or two Fourier components in the Hamiltonian, but if we wish to extend the method to include thousands of values of $n$ and a high-dimensional Hilbert space, truncating $H_{\rm F}$ is no longer viable. Fortunately, for the problem we shall be interested in (a radical pair in a weak RF magnetic field), the time-dependent interactions are orders of magnitude smaller than the static Hamiltonian $\hat{H}^{(0)}$, so we can use a perturbative method for the diagonalisation.

Consider for simplicity the case of a single applied Fourier component
\begin{equation}
 \hat H(t) = \hat H^{(0)} + \hat H'\cos(\omega t+\delta).
\end{equation}
This form of the Hamiltonian makes $H_{\rm F}$ block tri-diagonal, with $\frac{1}{2}H'e^{\pm i\delta}$ in the super- and sub-diagonal blocks. If the Hilbert space basis is the eigenbasis of $\hat{H}^{(0)}$, it is clear that there will be a degeneracy between a state in the $n^{\rm th}$ diagonal block of $H_{\rm F}$ and another in the $(n+1)^{\rm th}$ block if the applied frequency $\omega$ is resonant with the energy difference between two unperturbed eigenvalues. If the oscillating field is sufficiently weak, its effect will be confined to these resonant interactions. We shall therefore assume this to be the case, and use degenerate perturbation theory to calculate the first order corrections to the states brought into resonance (or near-resonance) by $\hat H'$.

As noted by Shirley,\cite{Shirley65} a near-resonance between two states, $E_\alpha + \omega \simeq E_\beta$, means there are nearly-degenerate two-dimensional sub-spaces involving state $\beta$ in diagonal block $n$ and state $\alpha$ in block $n+1$. The periodicity of $H_{\rm F}$ implies that there is a copy of the same $2\times 2$ sub-matrix between all pairs of adjacent diagonal blocks
\begin{equation}
  \tilde H_{\rm F} = \left(\begin{array}{cc}
 E_\alpha+\omega &\frac{1}{2}H'_{\alpha\beta}e^{+i\delta} \\
 \frac{1}{2}H'_{\beta\alpha}e^{-i\delta} &E_\beta
\end{array}\right),
\end{equation}
shifted by an integer multiple of $\omega I$.\\
\indent Diagonalising this $2\times2$ sub-matrix with the unitary matrix
\begin{equation}
  U = \left(\begin{array}{cc}
  U_{\alpha\alpha} &U_{\alpha\beta} \\
  U_{\beta\alpha} &U_{\beta\beta} \\
\end{array}\right)
\end{equation}
gives the first order corrected energies for states $\alpha$ and $\beta$ as the eigenvalues $\tilde E_\alpha + \omega$ and $\tilde E_\beta$. Since the multiple of $\omega I$ simply shifts eigenvalues, the altered energies and eigenstates can be written as
\begin{equation}
\begin{aligned}
 &\varepsilon_{\alpha n} = \tilde E_\alpha + n\omega, \ \ \ \ket{\varepsilon_{\alpha n}} = \ket{\alpha n}U_{\alpha\alpha} + \ket{\beta\,n-1}U_{\beta\alpha}\\
 &\varepsilon_{\beta n} = \tilde E_\beta + n\omega, \ \ \ \ket{\varepsilon_{\beta n}} = \ket{\beta n}U_{\beta\beta} + \ket{\alpha\,n+1}U_{\alpha\beta}.
\end{aligned}
\end{equation}
These approximate eigenvalues and eigenstates satisfy the periodicity conditions in Eqs.~(5) and~(6), as required. The other (non-resonant) energies and eigenstates are unaltered, $\varepsilon_{\gamma n} = E_\gamma+n\omega$ and $\ket{\varepsilon_{\gamma n}} = \ket{\gamma n}$, and we now have an approximate eigendecomposition of $H_{\rm F}$ to first order in $H'$. Clearly, this approximation rests on the assumption that $\omega\gg |H'|$, because if this were not the case the $2\times 2$ matrix in Eq.~(23) would not be isolated from other interactions.

\subsection{General procedure}

The method described above can be extended to treat more complicated situations -- including degeneracies in the unperturbed system, overlapping resonances, and a larger number of Fourier components, provided again that the spacing $\omega$ between these Fourier components is large compared with the strength of the perturbation. The general procedure is to build and diagonalise sub-matrices including as many as possible of the resonances and the appropriate coupling terms to obtain the first order energies. If there are multiple nearly-degenerate states, the dimension of the corresponding sub-space is enlarged to include them. It should be apparent that each state can only appear in one nearly-degenerate sub-matrix, all of the states with which is it nearly degenerate being by extension close in energy to one another.

One can construct examples to show that it may not always be possible to include {\em all} near resonances  without allowing the size of the nearly-degenerate sub-matrix to exceed the size of the Hilbert space. However, if the Hilbert space contains $N$ states, it is always possible to build an $N\times N$ nearly-degenerate sub-matrix of $H_{\rm F}$ that contains one copy of each Hilbert space state and captures (at least) the $N-1$ {\em closest} resonances (and typically many more). Assuming that the off-diagonal coupling terms in the Floquet Hamiltonian all have similar orders of magnitude, it is these closest resonances that will have the largest effect on the perturbed eigenvalues and eigenstates. Neglecting more distant  near resonances is clearly an approximation, but it is a convenient one to make, because diagonalising a Hilbert space-sized sub-matrix of $H_{\rm F}$ is no more expensive than diagonalising $H^{(0)}$. Since this procedure is also easily automated, it is the one we have adopted in the example calculations reported in Sec.~V.

\subsection{Computing the trace}

Having found an approximate eigendecomposition of $H_{\rm F}$, the final stage is to evaluate the expression for $\left<A(t)\right>$ in Eq.~(18) involving a double sum over the Floquet eigenstates. The number of these eigenstates is infinite, but only a finite number of them contribute to the double sum because of the structure of $\hat{A}_{\rm F}$ in Eq.~(12). For example, in the simple case where the approximate Floquet eigenstates are those in Eq.~(25), the matrix element $\bra{\varepsilon_{\alpha m}}\hat{A}_{\rm F}\ket{\varepsilon_{\beta n}}$ in Eq.~(18) is only non-zero when $m$ is 0 or 1 and $n$ is 0 or $-1$. The infinite sums over $n$ and $m$ therefore each collapse to just two terms, and an analogous simplification is obtained in the more general case in which a Hilbert space-sized nearly-degenerate sub-matrix of $H_{\rm F}$ is diagonalised to obtain the approximate Floquet eigenstates.

\section{Application to radical pairs}

With this Floquet machinery in hand, let us now return to the problem of simulating the spin dynamics of a RP subject to a RF magnetic perturbation.

\subsection{Singlet probability}

\begin{figure*}[t]
\includegraphics[scale=0.39]{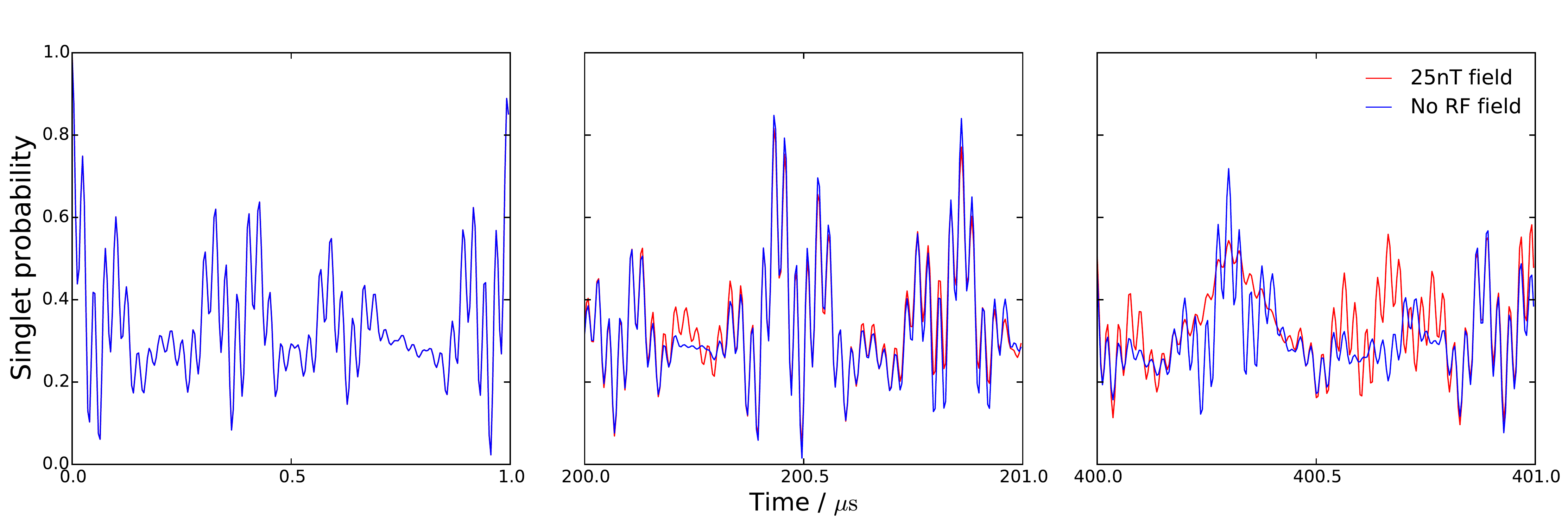}
\caption{Singlet probabilities $P_{\rm S}(t)$ for a model radical pair with and without a single resonant applied RF field. There are rapid (sub 25 ns) oscillations in both signals, but it takes longer than $100\ \mu{\rm s}$ for the effect of the RF field to become apparent.}
\end{figure*}

In the absence of any dipolar or exchange coupling between the two electrons, the unperturbed Hamiltonian that governs the spin evolution of a RP in the presence of a static external Zeeman field is\cite{Cintolesi03}
\begin{equation}
 \hat H^{(0)} = \hat H_1^{(0)} + \hat H_2^{(0)},
\end{equation}
where the individual radical Hamiltonians are
\begin{equation}
 \hat H_i^{(0)} = -\gamma_i\elecspin{i}{\cdot} {\bf B}^{(0)} + \sum_{k=1}^{N_i} \elecspin{i} {\cdot} {\bf A}_{ik} {\cdot} \hat{\bf I}_{ik}.
\end{equation}
Here the first term, in which $\gamma_i$ is the gyromagnetic ratio of the electron in radical $i$, is the Zeeman interaction of the electron spin $\hat {\bf S}_i$ with the static magnetic field ${\bf B}^{(0)}$. The second term contains the anisotropic hyperfine interactions between the electron spin and the $N_i$ nuclear spins $\hat {\bf I}_{ik}$ in the radical. We have neglected the comparatively weak Zeeman interactions of the nuclear spins with the magnetic field and the nuclear spin-spin coupling, but these could easily be added to Eq.~(27) if necessary.

In the presence of a radiofrequency magnetic field, the full Hamiltonian $\hat{H}(t)$ that governs the evolution of the radical pair has the form in Eq.~(1), where $\hat{H}^{(0)}$ is given in Eq.~(26) and when $n\not=0$
\begin{equation}
\hat{H}^{(n)} = \hat{H}^{(n)}_1+\hat{H}^{(n)}_2,
\end{equation}
with
\begin{equation}
\hat{H}^{(n)}_i = -\gamma_i\elecspin{i}\cdot {\bf B}^{(n)}e^{+i\delta_n}. 
\end{equation}
Here ${\bf B}^{(n)}$ is the (real) magnetic field vector of the $e^{+in\omega t}$ Fourier component of the RF radiation, and $\delta_n$ is a phase factor analogous to the $\delta$ in Eq.~(22). Note that the hermicity of $\hat{H}(t)$ demands that $\hat{H}^{(-n)}=\hat{H}^{(n)\dagger}$,  which implies that ${\bf B}^{(-n)}={\bf B}^{(n)}$ and $\delta_{-n}=-\delta_n$. 

The RP will typically be produced by a photo-induced electron transfer reaction from a singlet ground electronic state with equilibrium nuclear spin states. Since spin is conserved in this process, the initial density operator is\cite{Manolopoulos13}
\begin{equation}
 \hat\rho(0) = \frac{1}{Z_1Z_2}\hat P_{\rm S},
\end{equation}
where $Z_i=\prod_{k=1}^{N_i}(2I_{ik}+1)$ is the number of nuclear spin states in radical $i$ and $\hat P_{\rm S}$ is the singlet projection operator
\begin{equation}
 \hat P_{\rm S} = \frac{1}{4} - \elecspin{1} \cdot \elecspin{2}.
\end{equation}

Although there are other detection operators one could consider, the most relevant for our purposes is simply
\begin{equation}
\hat{A}=\hat{P}_{\rm S}.
\end{equation}
The corresponding time-dependent observable $\left<A(t)\right>$ is the probability $P_{\rm S}(t)$ that the RP is still in the singlet state at time $t$ after the initial photo-excitation. According to Eq.~(18), this is given by
\begin{equation}
\begin{aligned}
P_{\rm S}(t) &= \sum_{\alpha\beta} \sum_{mn} \bra{\varepsilon_{\alpha m}}\hat A_{\rm F}\ket{\varepsilon_{\beta n}}
\bra{\varepsilon_{\beta n}}\hat \rho_{\rm F}(0)\ket{\varepsilon_{\alpha m}}\\
&\times e^{+i (\varepsilon_{\alpha m} - \varepsilon_{\beta n})t},
\end{aligned}
\end{equation}
where from Eqs.~(12) and~(32),
\begin{equation}
\hat{A}_{\rm F} = \sum_{\alpha\beta} \ket{\alpha 0}\bra{\alpha}\hat{P}_{\rm S}\ket{\beta}\bra{\beta 0},
\end{equation}
and from Eqs.~(17) and (30),
\begin{equation}
\hat{\rho}_{\rm F}(0) = {1\over Z_1Z_2} \sum_{\alpha\beta}\sum_{mn} \ket{\alpha m}\bra{\alpha}\hat{P}_{\rm S}\ket{\beta}\bra{\beta n}.
\end{equation}
Note that Eq.~(33) ignores the possibility of electron spin relaxation, which would be difficult to include in the present formulation. The effect of electron spin relaxation on the FAD-tryptophan radical pair in cryptochrome has recently been investigated in a separate publication.\cite{Worster16}

Figure 2 shows the singlet probability as a function of time for a model RP (details given in Sec.~V) with and without a single $|{\bf B}^{(1)}|=25$ nT resonant RF perturbation applied. The plot highlights the oscillatory nature of $P_{\rm S}(t)$, and also confirms the assertion of Gauger \emph{et al.}\cite{Gauger11} that a very long ($>100\ \mu{\rm s}$) RP spin coherence is required for there to be a significant RF field effect with such a weak perturbation. The fact that this is many orders of magnitude longer than the period of the rapid oscillations in $P_{\rm S}(t)$ will be used to simplify the calculation of the singlet yield of the reaction below.

\subsection{Singlet yield}

In the context of avian magnetoreception, the observable of interest is the dependence of the singlet (or equivalently, since they sum to one, the triplet) yield of the radical pair recombination reaction on the direction of the static magnetic field.\cite{Schulten78} Assuming for simplicity that the recombination is symmetric ($k_{\rm S}=k_{\rm T}=k$ in Fig.~1), this singlet yield is given by\cite{Timmel98}
\begin{equation}
  \Phi_{\rm S} = k\int_0^{\infty}P_{\rm S}(t)e^{-kt} {\rm d}t.
\end{equation}
Inserting the expression for $P_{\rm S}(t)$ in Eq.~(33) and doing the integral over $t$ gives 
\begin{equation}
\begin{aligned}
\Phi_{\rm S} &= \sum_{\alpha\beta} \sum_{mn} \bra{\varepsilon_{\alpha m}}\hat A_{\rm F}\ket{\varepsilon_{\beta n}}
\bra{\varepsilon_{\beta n}}\hat \rho_{\rm F}(0)\ket{\varepsilon_{\alpha m}}\\
&\times {k^2\over k^2+(\varepsilon_{\alpha m} - \varepsilon_{\beta n})^2}.
\end{aligned}
\end{equation}

The form of this last equation reveals a simplification that is crucial for dealing with the infinite Floquet space. The factor of $k^2/[k^2+(\varepsilon_{\alpha m} - \varepsilon_{\beta n})^2]$ suppresses  contributions to the double sum in Eq.~(37) whenever the energy difference $|\varepsilon_{\alpha m}-\varepsilon_{\beta n}|$ is much larger than the recombination rate constant $k$: the coherent oscillations that are fast on the timescale of the RP lifetime have a negligible effect on the singlet yield. This implies that we can discard off-diagonal terms between states that differ widely in energy. In the case of long-lived radical pairs, the majority of terms in the double sum can be discarded. Indeed, when $k$ is sufficiently small, the only terms that contribute come from nearly-degenerate sub-spaces of the Floquet space -- exactly the same sub-spaces that we encountered when diagonalising $H_{\rm F}$ by degenerate perturbation theory in Sec.~III. 

\subsection{An alternative detection operator}

Some manipulation of the Floquet space detection operator will help us to exploit this simplification in the evaluation of $\Phi_{\rm S}$. Using the periodicity properties of the eigenvalues and eigenstates of $H_{\rm F}$ in Eqs.~(5) and~(6), one finds that Eq.~(10) can be re-written as
\begin{equation}
\begin{aligned}
\langle A(t;t_0)\rangle &= \sum_{\alpha\beta\gamma\delta}\sum_{mn} A_{\alpha\beta}
\bra{\beta q}e^{-i\hat H_{\rm F}(t-t_0)}\ket{\gamma m}e^{-im\omega t_0}\\
&\times \rho_{\gamma\delta}(t_0)e^{+in\omega t_0}\bra{\delta n}
e^{+i\hat H_{\rm F}(t-t_0)}\ket{\alpha q}
\end{aligned}
\end{equation}
for any integer $q$. Averaging this $q$ between $-Q$ and $Q$ and then taking the limit as $Q\to\infty$, it follows that the Floquet space detection operator in Eq.~(12) can be written equivalently as
\begin{equation}
\hat A_{\rm F} = \lim_{Q \rightarrow \infty} \frac{1}{2Q+1}\sum_{q=-Q}^{Q}\sum_{\alpha\beta} \ket{\alpha q}A_{\alpha\beta}\bra{\beta q}.
\end{equation}
In matrix form, the detection operator now has the Hilbert space-sized matrix $A$ in \emph{all} diagonal blocks
\begin{equation}
 A_{\rm F} = \lim_{Q \rightarrow \infty} \frac{1}{2Q+1}\left(\begin{array}{c|c|c|c|c}
... & & & &... \\ \hline
 &A &0 &0 & \\ \hline
 &0 &A &0 & \\ \hline
 &0 &0 &A & \\ \hline
... & & & &... \\
\end{array}\right).
\end{equation}

\subsection{Computational procedure}

In Sec.~III, we argued that the $N-1$ nearest degeneracies among the Floquet states can be condensed into an $N\times N$ Hilbert space-sized matrix $\tilde{H}_{\rm F}$ that is infinitely repeated, each copy differing only by a factor of $n\omega  I$ for Fourier index $n$. Thanks to the symmetry employed in constructing the alternative form of the Floquet space detection operator in Eq.~(40), both it and the initial density matrix are invariant to a change of Fourier index. It therefore follows that the corresponding density ($\tilde \rho_{\rm F}(0)$) and detection ($\tilde A_{\rm F}$) matrices for each copy of $\tilde H_{\rm F}$ are identical. To build these matrices, we simply insert the matrix elements from $A_{\rm F}$ and $\rho_{\rm F}(0)$ corresponding to those in $\tilde H_{\rm F}$.

As outlined above, in the long RP lifetime regime, we can approximately evaluate the double sum in the expression for $\Phi_{\rm S}$ by only including contributions from nearly-degenerate states. Indeed if $k\ll \omega$, we need only be concerned with terms that arise from the near-degeneracies that are captured in each Fourier-shifted copy of $\tilde{H}_{\rm F}$. To include these terms, it suffices to diagonalise a single copy of $\tilde H_{\rm F}$ with the Fourier index $n=0$, transform a single copy of $\tilde A_{\rm F}$ and $\tilde \rho_{\rm F}(0)$ into the eigenbasis (the eigenvectors of each effective Hamiltonian matrix shifted by $n\omega I$ being the same), and calculate the corresponding contributions to $\Phi_{\rm S}$.

Furthermore, because the expression for $\Phi_{\rm S}$ only depends on energy \emph{differences}, and each copy of $\tilde H_{\rm F}$ is merely shifted in energy by $n\omega$, the contribution to $\Phi_{\rm S}$ from each nearly-degenerate sub-space with a different Fourier index will be the same. Therefore, we have exactly $2Q+1$ identical contributions to the singlet yield. This factor cancels with the normalisation constant in the detection operator to leave a double sum over the states in the Hilbert space,
\begin{equation}
\Phi_{\rm S} \simeq \sum_{\alpha\beta=1}^N \left(\tilde{A}_{\rm F}\right)_{\alpha\beta}\Bigl(\tilde{\rho}_{\rm F}(0)\Bigr)_{\beta\alpha}\,{k^2\over k^2+(\tilde{E}_{\alpha}-\tilde{E}_{\beta})^2}.
\end{equation}
Here $\tilde{A}_{\rm F}$ is calculated without the factor of $1/(2Q+1)$ in Eq.~(40), and $\tilde{E}_{\alpha}$ and $\tilde{E}_{\beta}$ could be eigenvalues of any copy of $\tilde{H}_{\rm F}$, for example that with Fourier index $n=0$.

\section{Example calculations}

In order to test the accuracy of this Floquet-based approximation, we have carried out calculations on a small model system which could also be simulated exactly using a time-dependent propagation method. The system is comprised of a single proton on each radical with parallel axial hyperfine interaction tensors $(A_{xx},A_{yy},A_{zz}) = (-0.0636,-0.0636,1.0812)$ {\rm mT} and $(-0.0989,-0.0989,1.7569)$ {\rm mT}. These are based on the largest hyperfine couplings in the FAD-tryptophan RP in cryptochrome,\cite{Lee14} and are therefore representative of the hyperfine interactions that arise in biological systems.

Given our interest in the magnetic compass sense of migratory birds, which is believed to be associated with the sensitivity of the singlet yield to the direction of an Earth-strength ($\sim 50\ \mu$T) static magnetic field, we used both Eq.~(41) and the exact time-dependent propagation method to compute $\Phi_{\rm S}(\theta)$ as a function of the static field direction ${\bf B}^{(0)}(\theta)=50\ \mu{\rm T}\times \left(\hat{\bf x}\sin(\theta) + \hat{\bf z}\cos(\theta)\right)$. This was done for a variety of radical pair recombination rate constants and both monochromatic and broadband RF radiation in order to assess the accuracy of Eq.~(41) in a variety of regimes.

\subsection{Monochromatic radiation}

We initially tested the Floquet method with a perturbation comprised of a single Fourier component
\begin{equation}
\hat{H}^{(1)}_ie^{+i\omega t} = -\gamma_i\hat{\bf S}_i\cdot {\bf B}^{(1)}e^{+i(\omega t+\delta_1)},
\end{equation}
and its adjoint
\begin{equation}
\hat{H}^{(-1)}_ie^{-i\omega t} = -\gamma_i\hat{\bf S}_i\cdot {\bf B}^{(1)}e^{-i(\omega t+\delta_1)}.
\end{equation}
Here $\gamma_i$ was taken to be the gyromagnetic ratio of a free electron ($\gamma_{\rm e}/2\pi=-0.028025$ MHz/$\mu$T) for each radical in the pair. We arbitrarily selected the frequency $\nu=\omega/2\pi = 2.0961$ MHz, which comes into resonance with a spacing between energy levels of $\hat{H}^{(0)}(\theta)$ near $\theta=40^{\circ}$ and $\theta=140^{\circ}$. We also chose an arbitrary initial phase $\delta_1$ and inclination $\hat{\bf B}^{(1)}$ of the RF magnetic field.

\begin{figure}[t]
\begin{center}
 \includegraphics[scale=0.425]{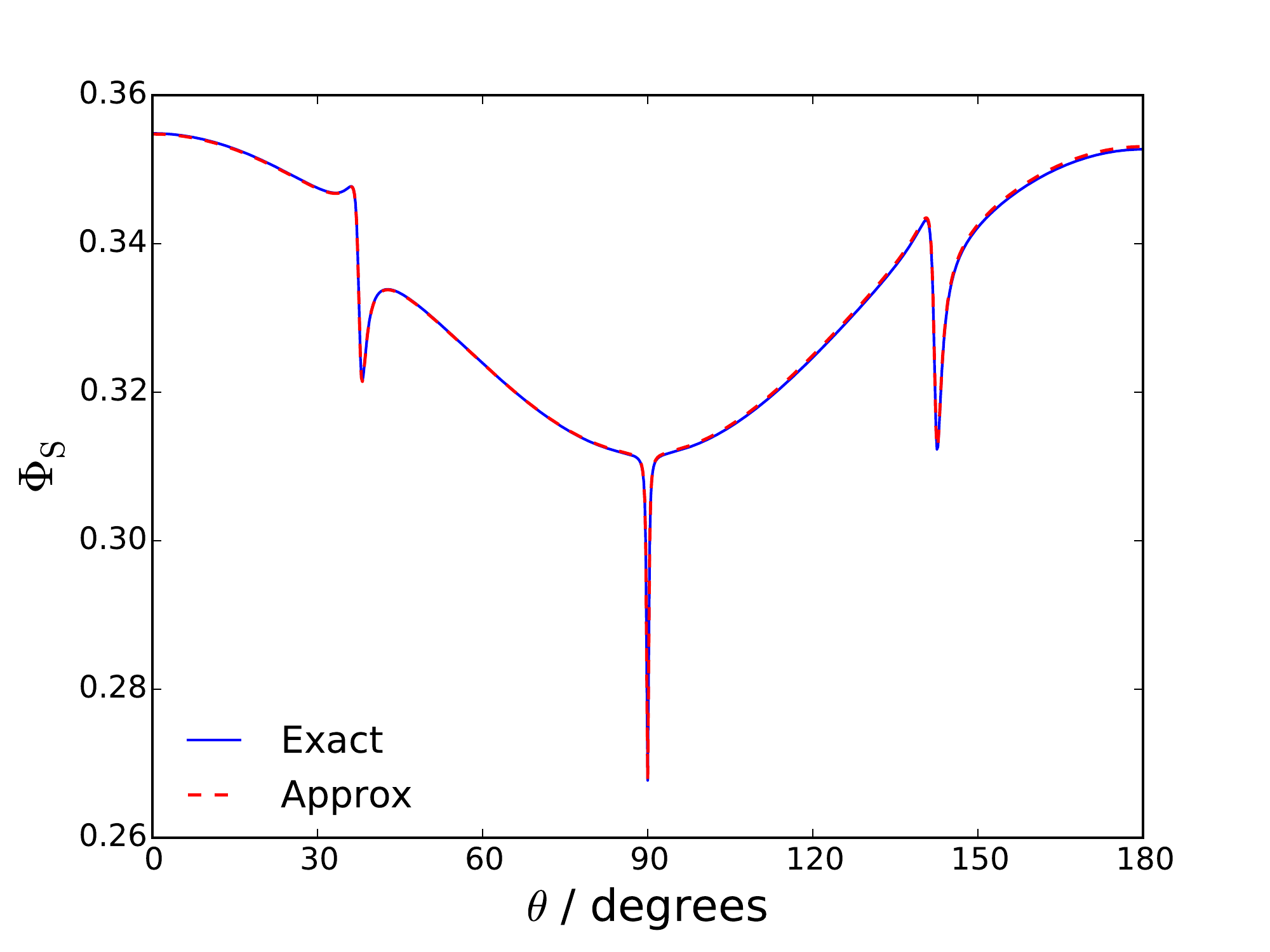}
\caption{Exact (time-dependent propagation) and approximate (Floquet perturbation theory) singlet yields for a simple model radical pair in the presence of a monochromatic RF magnetic field, as a function of the angle $\theta$ between the static magnetic field axis and the RP $z$ axis. Here the RF magnetic field strength is $|{\bf B}^{(1)}|=250$ nT and the lifetime of the radical pair is $\tau=1/k = 100\ \mu$s.}
\end{center}
\end{figure}

Figure~3 compares the singlet yield obtained from Eq.~(41) with the exact quantum mechanical singlet yield as a function of $\theta$ for this model problem, with a perturbation strength of $|{\bf B}^{(1)}|=250$ nT and a radical pair lifetime of $\tau=1/k=100\ \mu$s. The two curves, which each consist of three spikes superimposed on a mildly varying sinusoidal background, are seen to be identical to graphical accuracy.

The central spike at $\theta=90^{\circ}$ in Fig.~3 has been discussed in detail in a recent paper.\cite{Hiscock16} It arises from a narrowly avoided crossing between the eigenvalues of $\hat{H}_0(\theta)$, which gives rise to a negative Lorentzian lineshape in $\Phi_{\rm S}(\theta)$. The narrowness of the spike has been suggested as a possible explanation for the high precision of the magnetic compass sense of migratory birds.\cite{Akesson01,Lefeldt15} 

The additional spikes at $\theta\simeq 40^{\circ}$ and $\theta\simeq 140^{\circ}$ in Fig.~3 are generated by the RF field: they occur when the frequency of this field becomes resonant with an energy level spacing of $\hat{H}^{(0)}(\theta)$. These spikes are not purely Lorentzian: they have more complicated lineshapes that are sensitive to the initial phase and inclination of the RF magnetic field.

\begin{figure}[t]
\begin{center}
\includegraphics[scale=0.425]{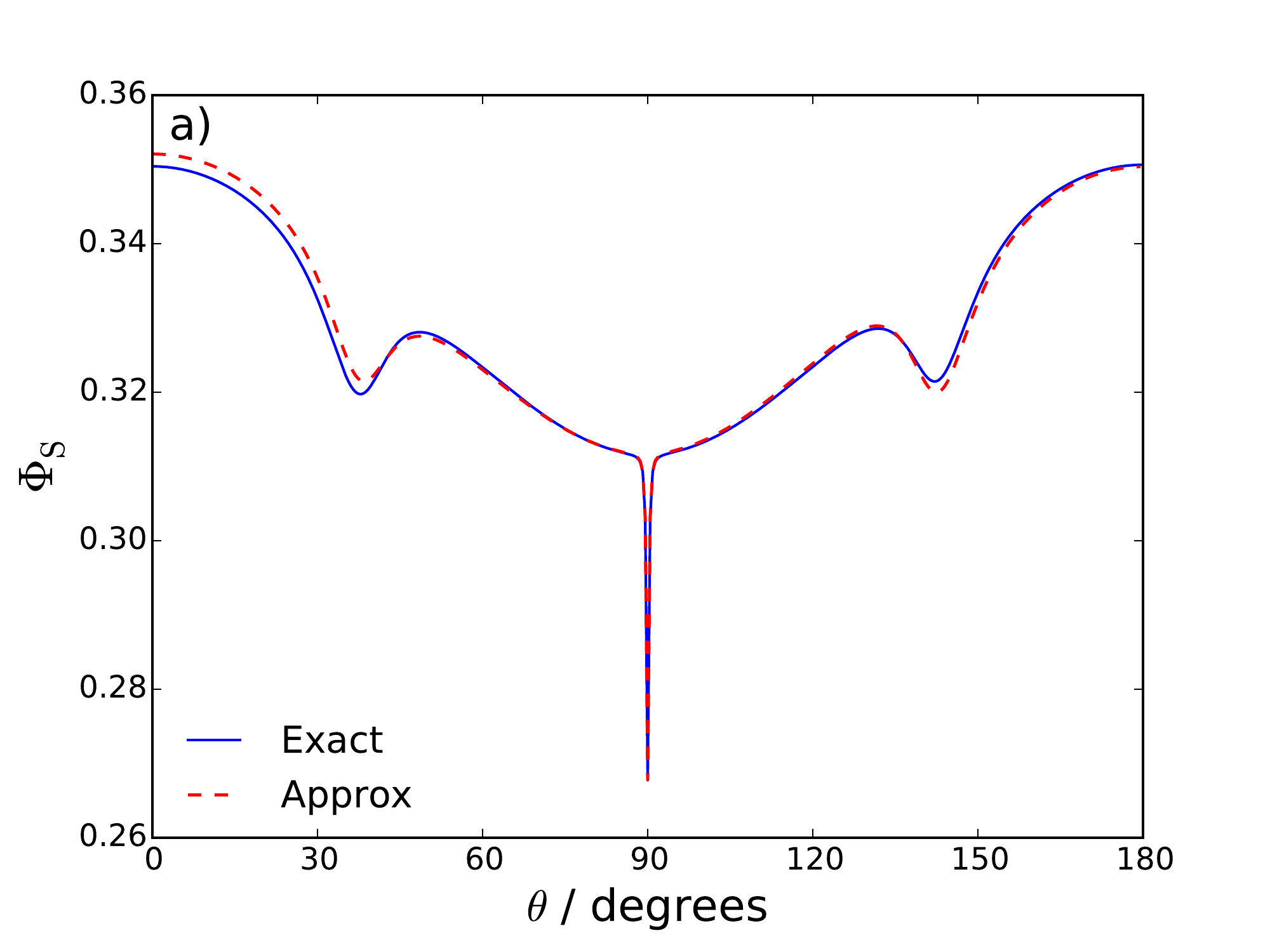}
\includegraphics[scale=0.425]{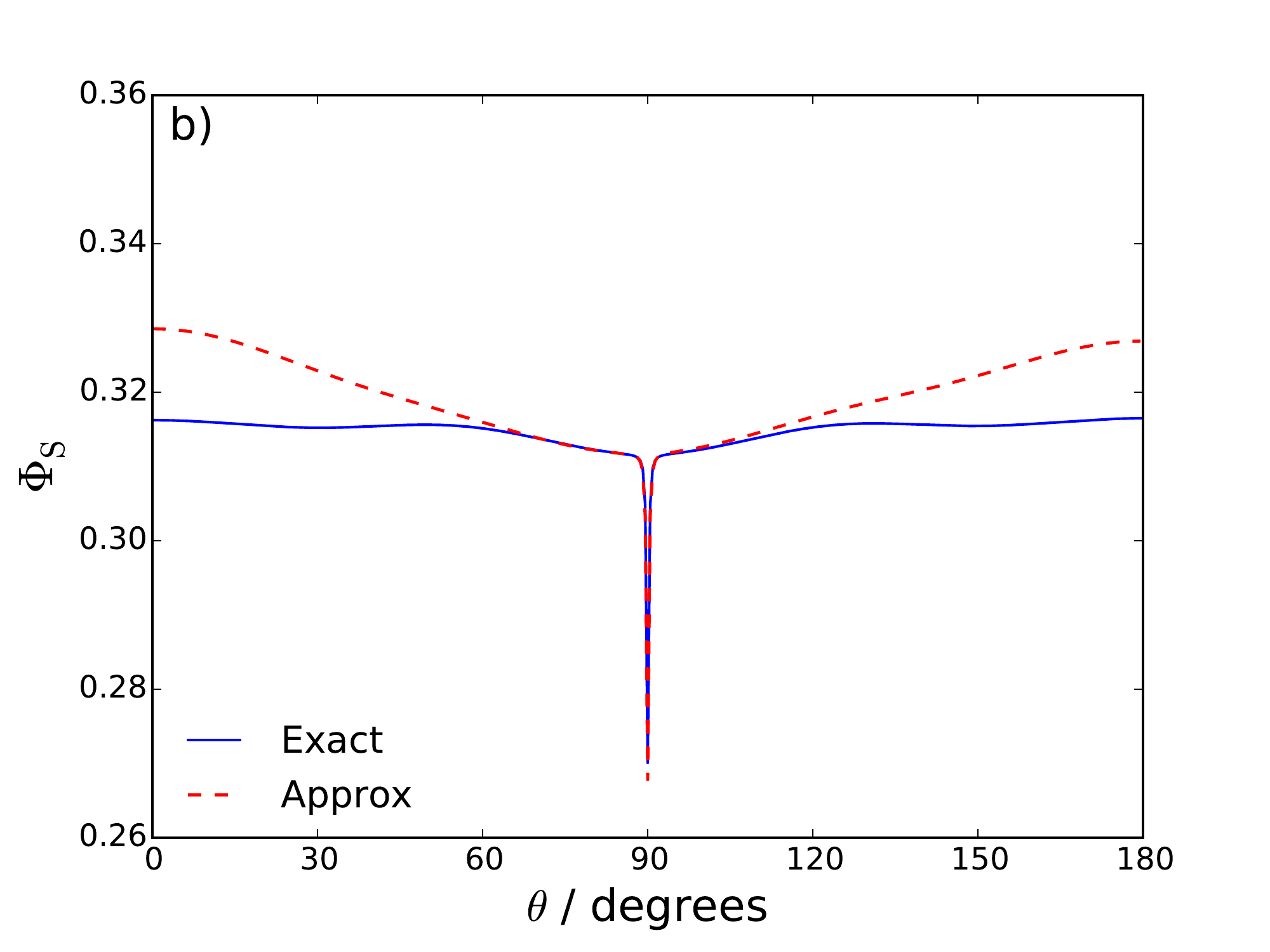}
\caption{As in Fig.~3, but for (a) $|{\bf B}^{(1)}|=2.5\ \mu$T and (b) $|{\bf B}^{(1)}|=25\ \mu$T.}
\end{center}
\end{figure}

The results in Fig.~3 show that the Floquet approximation is essentially exact for this model problem when $|{\bf B}^{(1)}| = 250$ nT and $\tau = 100\ \mu$s. But from the derivation presented in Secs.~III and~IV, one would expect it to become less accurate as the strength of the perturbation is increased and the radical pair lifetime is decreased. In order to explore this, we have gone on to investigate what happens when $|{\bf B}^{(1)}|$ is increased and $\tau$ is decreased by 1-2 orders of magnitude. 

The effect of increasing $|{\bf B}^{(1)}|$ is shown in Fig.~4. As the strength of the RF perturbation is increased, the associated resonances at $\theta\simeq 40^{\circ}$ and $\theta\simeq 140^{\circ}$ broaden, while the central spike at $\theta=90^{\circ}$ remains unchanged. The broadening of the resonances is captured reasonably well by the Floquet approximation up to a magnetic field strength of $|{\bf B}^{(1)}| = 2.5\ \mu$T. But by the time $|{\bf B}^{(1)}| = 25\ \mu$T, the approximation has broken down. Since this is already half the strength of the static Zeeman field, $|{\bf B}^{(0)}(\theta)|=50\ \mu$T, it is not surprising that it is no longer valid to treat it as a perturbation. 

The effect of decreasing $\tau$ is shown in Fig.~5. This effect is qualitatively different from that of increasing $|{\bf B}^{(1)}|$, in that it decreases the amplitudes of both the RF field-induced resonances at $\theta\simeq 40^{\circ}$ and $\theta\simeq 140^{\circ}$ and the central spike at $\theta=90^{\circ}$. The narrow angular features in the singlet yield are all washed out by the lifetime broadening associated with a shorter-lived RP. This effect is seen to be captured almost quantitatively by the Floquet approximation when $\tau=10\ \mu$s, but not for a RP lifetime as short as 1 $\mu$s.

\begin{figure}[t]
\begin{center}
\includegraphics[scale=0.425]{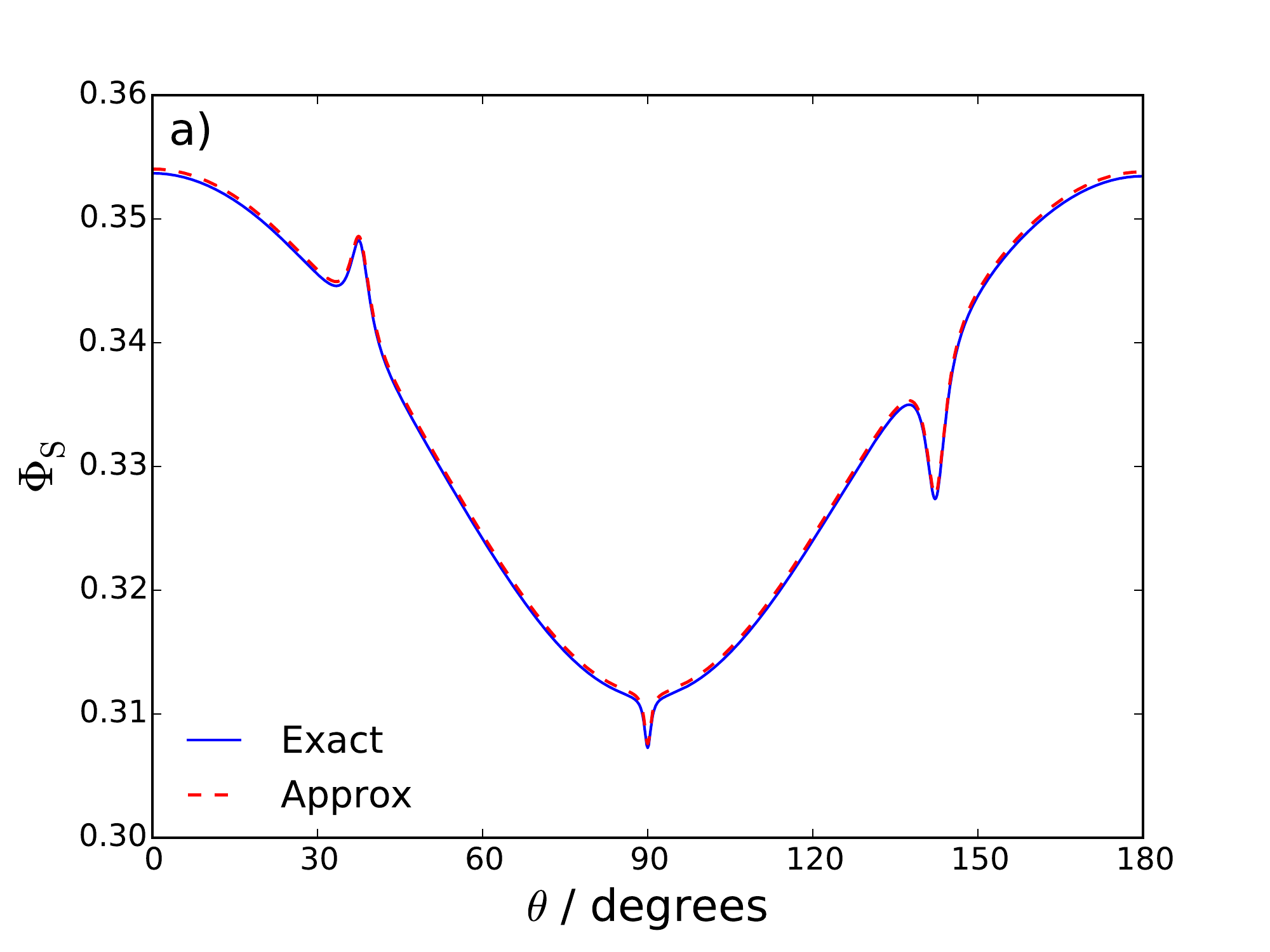}
\includegraphics[scale=0.425]{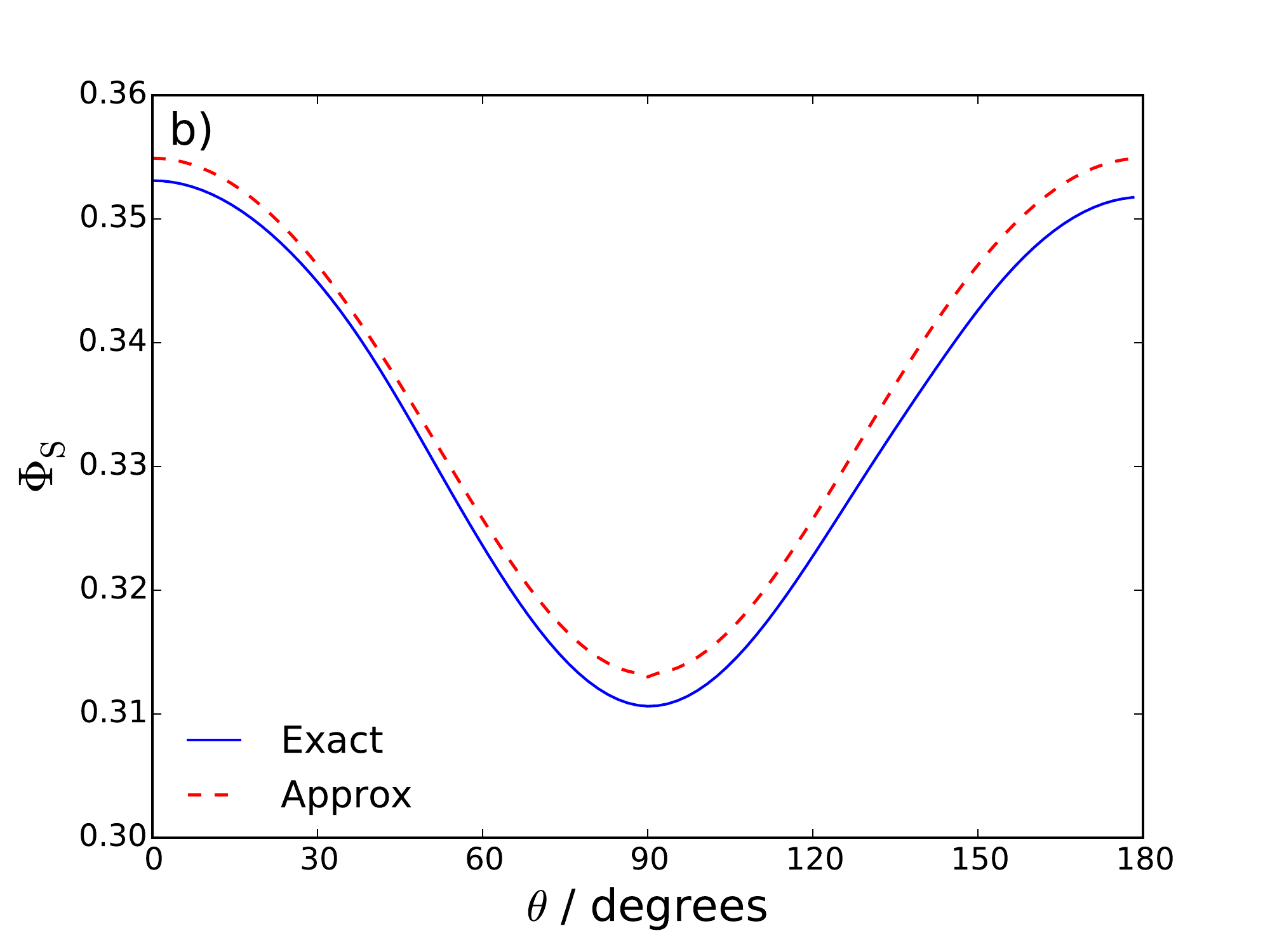}
\caption{As in Fig.~3, but for (a) $\tau=10\ \mu$s and (b) $\tau=1\ \mu$s.}
\end{center}
\end{figure}

\subsection{Broadband noise}

We have also tested the Floquet approximation for a broadband RF perturbation, made up of equally spaced Fourier components.\cite{Comb} For this, we used a base frequency of $\nu=\omega/2\pi=15.9$ kHz, including Fourier components with $|n|$ between 100 and 2000. The band limits were therefore 1.59 and 31.8 MHz. Given the much larger number of Fourier components in this RF field compared with the monochromatic example considered above, the strength of each component was reduced to $|{\bf B}^{(n)}|=25$ nT to give a manageable perturbation. The phase $\delta_n$ and inclination $\hat{\bf B}^{(n)}$ of each component were again chosen randomly, and the calculations were performed with a radical pair lifetime of $\tau=1$ ms.

\begin{figure}[t]
\begin{center}
\includegraphics[scale=0.45]{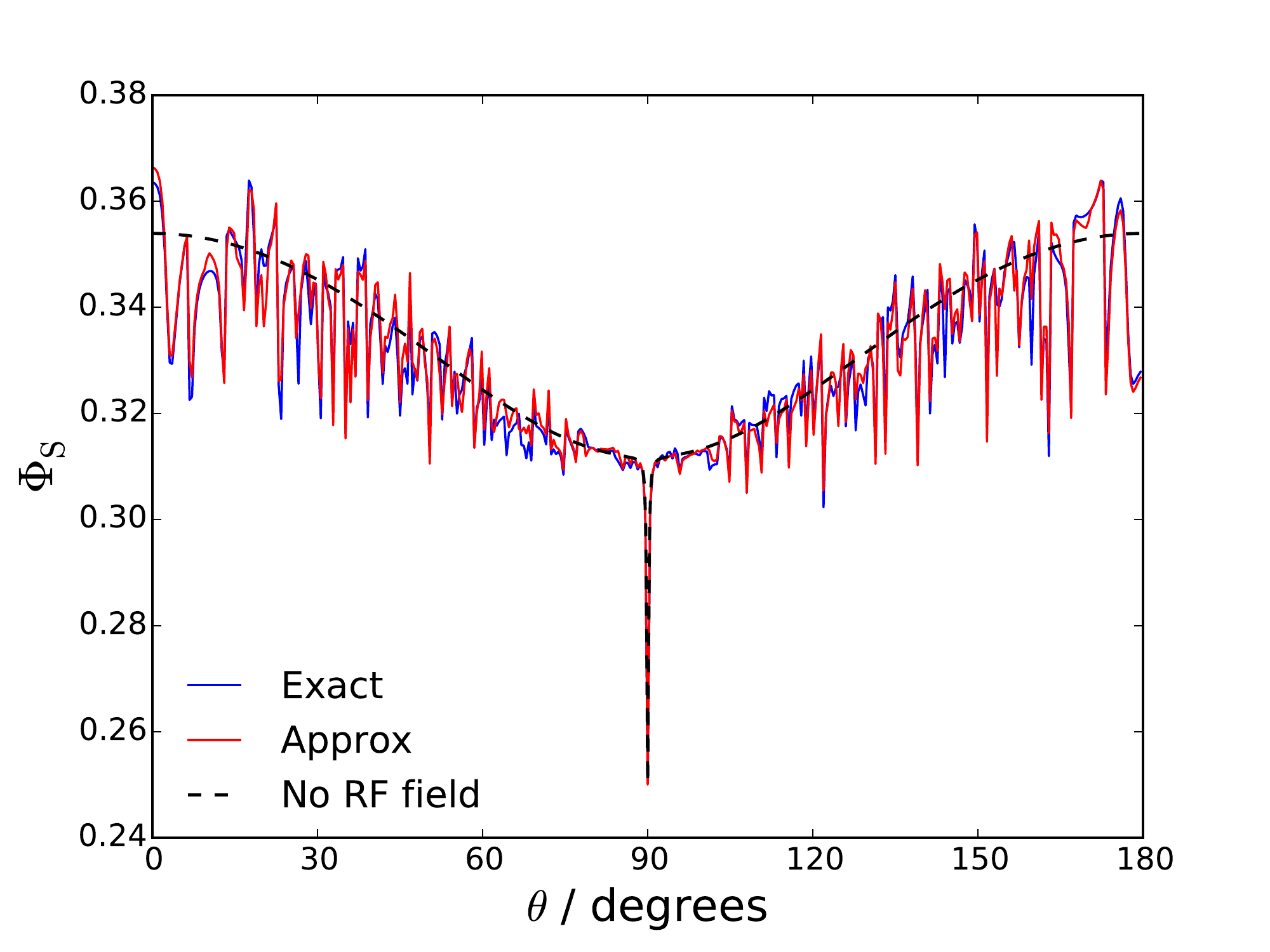}
\caption{As in Fig.~3, but with the monochromatic RF field replaced by a broadband noise field containing 1901 equally-spaced frequencies between 1.59 and 31.8 MHz. The exact results were again obtained using a time-dependent propagation method, and the approximate results using the Floquet perturbation theory expression in Eq.~(41). In this case, the singlet yield in the absence of the RF field is also shown for comparison.}
\end{center}
\end{figure}

Figure 6 shows the $\Phi_{\rm S}(\theta)$ curves for this broadband noise model as calculated with an exact time-dependent propagation method and with the Floquet formula in Eq.~(41). The singlet yield in the absence of the RF field is also shown for comparison.  Although there are now many more resonances than in the monochromatic case, one sees that the agreement between the exact and Floquet results is again very good, both in terms of the positions of the RF field-induced resonances and their intensities relative to the RF field-free signal.  

Finally, to test whether the Floquet approximation remains reliable for larger radical pairs, we have repeated this broadband noise calculation with an additional proton added to each radical. The additional proton in the first radical was given the hyperfine tensor 
\begin{equation*}
\begin{pmatrix}
A_{xx} & A_{xy} & A_{xz} \cr A_{yx} & A_{yy} & A_{yz} \cr A_{zx} & A_{zy} & A_{zz} \cr
\end{pmatrix}
=
\begin{pmatrix}
-0.9920 & -0.2091 & -0.2003 \cr -0.2091 & -0.2631 & +0.2803 \cr -0.2003 & +0.2803 & -0.5398 \cr
\end{pmatrix},
\end{equation*}
and that in the second radical the axial hyperfine tensor $(A_{xx},A_{yy},A_{zz}) = (-0.0190,-0.0190,1.7569)$, both in mT. These are again based on the hyperfine interactions of magnetic nuclei in the FAD-tryptophan radical pair in cryptochrome.\cite{Lee14} All the other details of the calculation were kept the same.

The results are shown in Fig.~7, again with the RF field-free signal included for comparison. Now this field-free signal contains three spikes due to avoided crossings between the energy levels of $\hat{H}^{(0)}(\theta)$, the original narrow spike at $\theta=90^{\circ}$ and two broader spikes at $\theta\simeq 70^{\circ}$ and $\theta\simeq 100^{\circ}$. The effect of the broadband RF field is seen to be much the same as before, giving rise to a dense forest of narrow resonances on top of the RF field-free background. 

The Floquet results in Fig.~7 are again in good agreement with the exact results in terms of the density of the resonances and their positions, although they are not in quite such good agreement as those in Fig.~6 in terms of all of the resonance intensities. This is slightly concerning for future applications of the Floquet method to larger and more realistic radical pairs, but one might at least hope that its predictions would be qualitatively reasonable. Since the exact (time-dependent propagation) results in Fig.~7 took five orders of magnitude more computer time to generate than the Floquet results, we believe that there really is no practical alternative to the Floquet method for studying long-lived radical pairs with many more nuclear spins in RF magnetic fields.

\begin{figure}[t]
\begin{center}
\includegraphics[scale=0.45]{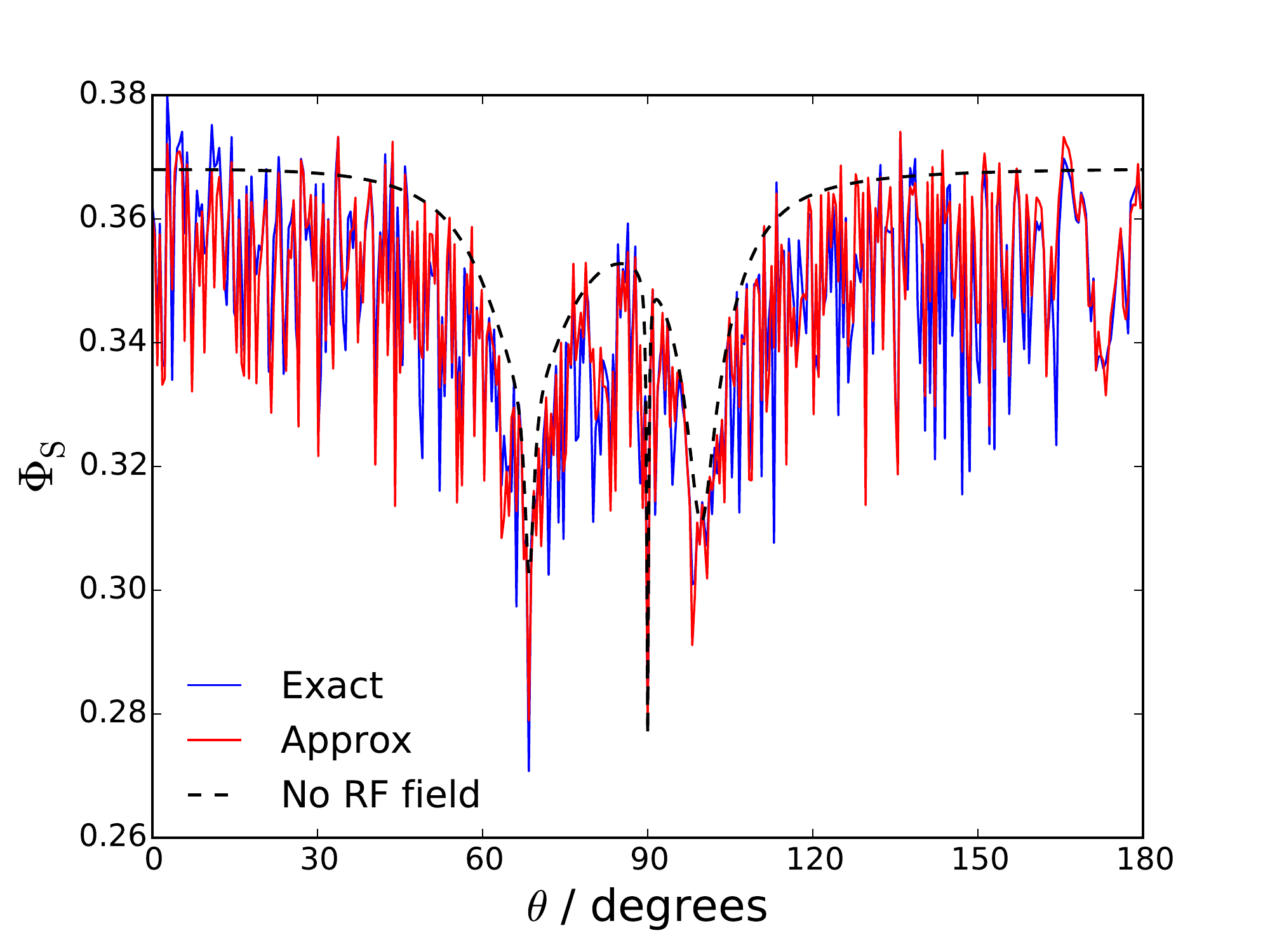}
\caption{As in Fig.~6, but for a larger radical pair with two hyperfine-coupled nuclear spins in each radical.}
\end{center}
\end{figure}

\section{Concluding Remarks}

In this paper, we have shown how a modified formulation of Floquet theory combined with degenerate perturbation theory can be used to provide a reasonable approximation to the singlet yield of a radical pair recombination reaction in the presence of a RF magnetic field. In example calculations of small radical pairs with biologically reasonable hyperfine couplings, the resulting Floquet approximation has been shown to be quantitatively accurate for monochromatic RF magnetic fields with $|{\bf B}^{(1)}|\le 2.5\ \mu$T and radical pair lifetimes $\tau\ge 10\ \mu$s (see Figs.~3 to 5), and qualitatively reasonable for a broadband magnetic field with $2\times 1901$ Fourier components each with $|{\bf B}^{(n)}|\le 25$ nT (see Figs.~6 and~7). 

Since our implementation of the Floquet approximation only involves matrix operations in which the matrices are the size of the Hilbert space [see Eq.~(41)], its computational cost is only a couple of times larger than that of a standard (RF field-free) radical pair singlet yield calculation done by Hamiltonian matrix diagonalisation. A large number of such calculations have been performed in recent years involving radical pairs with as many as twenty or so nuclear spins,\cite{Cintolesi03,Lee14,Cai10,Cai11,Lau12} typically by exploiting the separability of the Hamiltonian in Eq.~(26). Any radical pair that has previously been studied in this way could clearly now be exposed to a RF magnetic field and its response studied with the help of the present Floquet approximation, including the FAD-tryptophan radical pair in cryptochrome that has been suggested as the origin of the magnetic compass sense of migratory birds.\cite{Schulten78,Hore16}

Insofar as the disruption of this compass sense by RF radiation is concerned, the present results in Figs.~6 and~7 are already rather interesting. If it is true, as suggested in Ref.~\onlinecite{Hiscock16}, that the precision of the compass arises from the narrowness of the spike at $\theta=90^{\circ}$ (which is present in the RF field-free signal), then one could imagine that the many additional narrow resonances that are generated by the (broadband) RF field might well distract the bird from the true north-south axis and lead to disorientation. However, we should stress that these calculations were performed using vastly over-simplified models of the FAD-tryptophan radical pair involving just 2 and 4 nuclear spins, and that the broadband noise used in the calculations had an intensity of $|{\bf B}^{(n)}|=25$ nT in each of its $2\times 1901$ Fourier components (giving a root-mean-square fluctuating magnetic field averaged over a cycle of 154 nT). The real FAD-tryptophan radical pair contains many more hyperfine-coupled nuclear spins, and experimental investigations of the disorientation of birds by broadband electromagnetic noise have involved rather weaker RF magnetic fields.\cite{Ritz04,Engels14,Schwarze16} Both of these factors (and also the role of electron spin relaxation\cite{Worster16}) will have to be taken into account before we can draw any firm conclusions about the effect of RF radiation on the avian compass. We plan to use the present Floquet theory to investigate the effect of weaker RF magnetic fields on more realistic models of the FAD-tryptophan radical pair in a future  article.

Finally, we should point out that we have simply followed earlier work\cite{Hore16} in assuming that the magnetic component of the RF radiation disrupts the bird's magnetic compass. This may or may not be the case.\cite{Hore16} It is also conceivable that the electric component of the radiation plays some role. In any event, we do believe that the theory we have developed here provides a practical way to study the effect of RF magnetic fields on a wide variety of radical pairs that arise in biological and also chemical\cite{Lersch89,Koptyug90,Batchelor92,Henbest04} systems.\\

\begin{acknowledgments}
We gratefully acknowledge funding from the European Research Council under the European Union's 7th Framework Programme, FP7/2007-2013/ERC Grant Agreement No.~340451, and from the US Air Force (USAF) Office of Scientific Research under the Air Force Materiel Command, USAF Award FA9550-14-1-0095.
\end{acknowledgments}


\begin{thebibliography}{99}
\bibitem{Comb}
{Strictly speaking, this is not broadband noise, but rather a broadband \lq\lq frequency comb". A RF signal with two or more incommensurate base frequencies would provide a better description of broadband noise. It is possible to generalise the present theory to deal with such a signal, by adopting a multimode version of Floquet theory of the sort described in Refs.~24 and~30. However, this generalisation breaks down whenever $n\omega_1+m\omega_2\simeq 0$, which can certainly happen when there are many Fourier components in each mode. For our target application, we feel that it is more important to include many Fourier components within a given RF band than to have a faithful description of true \lq\lq noise", and for this a frequency comb is ideal.}
\end{thebibliography}
\end{document}